\documentclass{article}
\usepackage{graphics,epsfig}
\title {The Description of Joint Quantum Entities \\ and the Formulation of a
Paradox\footnote{Published as: Aerts, D., 2000, ``The description of joint quantum entities and the formulation of a
paradox", {\it International Journal of Theoretical Physics}, {\bf 39}, 483-496.}}
\author {Diederik Aerts\thanks{FUND and CLEA, Brussels Free University, Pleinlaan 2, 1050
Brussels; e-mail: diraerts@vub.ac.be}}
\date {}
\font\smallroman=cmr10 at 8pt

\newtheorem{theorem}{Theorem}

\newtheorem{sqmprinciple}{SQM Principle}
\newtheorem{physprinciple}{Physical Principle}
 
\newtheorem{cqmprinciple}{CQM Principle}
\begin{document}

\maketitle

\begin{abstract}
\noindent We formulate a paradox in relation to the description of a joint entity
consisting of two subentities by standard quantum mechanics. We put forward a proposal
for a possible solution, entailing the interpretation of `density states' as `pure
states'. We explain where the inspiration for this proposal comes from and how its validity can be
tested experimentally. We discuss the consequences on quantum axiomatics of the proposal.
\end{abstract}

\section{Formulation of the Paradox} \label{sec01}
Quantum mechanics, after more than 70 years, still poses fundamental problems of
understanding. Many physicists believe these problems are `only'
problems of `physical interpretation' of the
mathematically well defined `standard formalism'. In this paper we will show that
this is not necessarily so. We will show that the problem of quantum mechanics connected to
the existence of nonproduct states in the situation of the description of the joint entity
of two quantum entities may well indicate that a change of the standard
formalism is necessary.

If we mention the `standard formalism' of quantum mechanics we mean the
formalism as exposed, for example, in the book of John von Neumann (1932), and we
will refer to it by SQM. Often the name `pure state' is used to indicate a
state represented by a ray of the Hilbert space. We will however use it in the physical sense:
a `pure state' of an entity represents the reality of this entity. As a consequence it is
natural to demand that an entity `exists' if and only if at any moment it is in one and only
one `pure state'. Let us formulate this as a principle, since it will play a major role in our
analysis. 
\begin{physprinciple} \label{physprinc01}
If we consider an entity, then this entity `exists' at a certain moment iff it `is'
in one and only one pure state at that moment. 
\end{physprinciple}
We denote pure states of an entity
$S$ by means of symbols $p, q, r, ...$ and the set of all pure states by $\Sigma$.
We
mention that in (Aerts, 1984a, 1999a), where aspects of the problem that we investigate
in the present paper are analyzed, a `pure state' is called a `state'.

A state represented by a ray of the Hilbert space will be called a `ray
state'. We denote rays by symbols
${\bar x}, {\bar y}, ...$, where $x, y, ... \in {\cal H}$ and we denote $p_x$ the `ray state'
represented by the ray ${\bar x}$. With each ray ${\bar
x}$, $x \in {\cal H}$, corresponds one and only one ray state $p_x$. One of the principles of
standard quantum mechanics is that `pure states' are `ray states'.
\begin{sqmprinciple} \label{qmprinc01}
Consider an entity $S$ described by SQM in a Hilbert space ${\cal H}$. Each ray state
$p_x$, $x \in {\cal H}$ is a pure state of $S$, and each pure state of $S$ is of this form.
\end{sqmprinciple}
The problem that we want to consider in this paper appears in the SQM description by
 of the joint entity $S$ of two quantum entities $S_1$ and $S_2$. 

\begin{sqmprinciple} \label{qmprinc02}
If we consider two quantum
entities
$S_1$ and
$S_2$ described by SQM in Hilbert spaces ${\cal H}_1$ and ${\cal
H}_2$, then the joint quantum entity
$S$ consisting of these two quantum entities is described by SQM in
the tensor product Hilbert space
${\cal H}_1 \otimes {\cal H}_2$. 
The subentities $S_1$ and $S_2$ are in
ray states $p_{x_1}$ and $p_{x_2}$, with $x_1 \in {\cal H}_1$ and $x_2 \in {\cal H}_2$,
iff the joint entity $S$ is in a ray state $p_{x_1 \otimes x_2}$. 
\end{sqmprinciple}
In relation to the situation
of a joint entity consisting of two subentities there is another physical principle we
generally imagine to be satisfied.
\begin{physprinciple} \label{physprinc02}
If an entity is the joint entity of two subentities then the entity exists at a certain
moment iff the subentities exist at that moment.
\end{physprinciple}
Let us introduce the concept of `nonproduct vectors' of the tensor product. For $z \in {\cal H}_1
\otimes {\cal H}_2$ we say that $z$ is a nonproduct vector iff $\not\exists \ z_1 \in {\cal H}_1,
z_2
\in {\cal H}_2: z = z_1 \otimes z_2$. We are now ready to formulate the theorem that
points out the paradox we want to bring forward.
\begin{theorem} Physical Principle~\ref{physprinc01}, Physical
Principle~\ref{physprinc02}, SQM Principle~\ref{qmprinc01} and SQM Principle \ref{qmprinc02} cannot
be satisfied together.
\end{theorem} \label{theor01}
Proof: Suppose the four principles
are satisfied. This leads to a contradiction. Consider the
joint entity $S$ of two subentities $S_1$ and $S_2$, described by SQM. From SQM Principle
\ref{qmprinc02} follows that if
$S_1$ and $S_2$ are described in Hilbert spaces
${\cal H}_1$ and ${\cal H}_2$ then $S$ is described in the Hilbert space ${\cal H}_1
\otimes {\cal H}_2$. Let us consider a nonproduct vector $z \in {\cal H}_1 \otimes {\cal
H}_2$ and the ray state $p_z$. From SQM Principle~\ref{qmprinc01} follows that
$p_z$ represents a pure state of $S$. Consider a moment where $S$ is in state $p_z$. Physical
Principle~\ref{physprinc01} implies that $S$ exists at that moment and from Physical Principle
\ref{physprinc02} we infer that $S_1$ and $S_2$ also exist at that moment. Physical
Principle
\ref{physprinc01} then implies that $S_1$ and $S_2$ are respectively in pure states $p_1$
and
$p_2$. From SQM Principle \ref{qmprinc01} it
follows that there are two rays ${\bar z_1}$ and ${\bar z_2}$, $z_1 \in {\cal H}_1$ and
$z_2 \in {\cal H}_2$ such that $p_1 = p_{z_1}$ and $p_2 = p_{z_2}$. From SQM
Principle~\ref{qmprinc02} then follows that
$S$ is in the state $p_{z_1 \otimes z_2}$ which is not $p_z$ since the ray generated by
$z_1
\otimes z_2$ is different from $\bar z$. Since both $p_z$ and $p_{z_1 \otimes z_2}$ are
pure states, this contradicts Physical Principle~\ref{physprinc01}.

\section{An Alternative Solution to the Paradox}

The fundamental problems of the SQM description of the joint entity of two subentities had already
been remarked a long time ago. The Einstein Podolsky Rosen paradox and later research on the
Bell inequalities are related to this difficulty (Einstein et al., 1935; Bell, 1964). It are indeed
states like
$p_z$, with $z$ a nonproduct vector, that give rise to the violation of the Bell
inequalities and that generate the typical EPR correlations between the subentities. Most of the
attention in this earlier analysis went to the `non local' character of these
EPR correlations. The states of type $p_z$ are now generally called
`entangled' states. The problem (paradox) related to entangled states that we have
outlined in section~\ref{sec01} has often been overlooked, although noticed and partly mentioned in
some texts (e.g. Van Fraassen, 1991, section 7.3 and references therein).

The problem of the description of a joint entity has also been studied within
axiomatic approaches to SQM. There it was shown that some of the axioms that are needed for SQM
are not satisfied for certain well defined situations of a joint entity consisting of two
subentities (Aerts, 1982, 1984a; Pulmannov\`a, 1983, 1985; Randall and Foulis, 1981). More
specifically, it has been shown in Aerts (1982) that the joint entity of two
separated entities cannot be described by SQM because of two axioms: weak modularity and the
covering law. This shortcoming of SQM is proven to be at the origin of the EPR
paradox (Aerts, 1984b, 1985a,b). It has also been shown that different formulations of the
product within the mathematical categories employed in the axiomatic structures do not coincide with
the tensor product of Hilbert spaces (Aerts, 1984a; Pulmannov\`a, 1983, 1985; Randall and
Foulis, 1981). Again certain axioms, orthocomplementation, covering law and atomicity, are causing
problems.

All these findings indicate that we are confronted with a deep problem that has
several complicated  and subtle aspects. A very extreme attitude was to consider entangled states
as artifacts of the mathematical formalism and hence not really existing in nature. Meanwhile,
entangled states are readily prepared in the laboratory and the corresponding EPR correlations have
been detected in a convincing way. This means that it is very plausible to acknowledge the
existence of entangled states as `pure states' of the joint entity in the sense of Physical
Principle \ref{physprinc01}.

As a result of earlier research we have always been inclined to believe
that we should drop Physical Principle \ref{physprinc02} to resolve the paradox (Aerts, 1984a,
1985a,b; see also Aerts, Broekaert and Smets, 1999). This comes to considering
the subentities
$S_1$ and
$S_2$ of the joint entity
$S$ as `not existing' if the joint entity is in an entangled state. We still believe that this
is a possible solution to the paradox and refer for example to Valckenborgh (1999) and Aerts and
Valckenborgh (1999) for further structural elaboration in this direction. In the present paper we
would like to bring forward an alternative solution.
To make it explicit we introduce the concept of `density state', which is a state
represented by a density operator of the Hilbert space. We denote density operators by symbols $W,
V, ...$ and the corresponding density states by
$p_W, p_V, ...$. With each density operator $W$ on
${\cal H}$ corresponds one and only one density state $p_W$. Within SQM density states are supposed
to represent mixtures, i.e. situations of lack of knowledge about the pure state. The way out
of the paradox we propose in the present paper consists of considering the density states as pure
states. Hence, in this sense, SQM Principle 1 would be false and substituted by a new
principle.

\begin{cqmprinciple} \label{cqmprinc}
Consider an entity $S$ described in a Hilbert space ${\cal H}$. Each density state
$p_W$, where $W$ is a density operator of ${\cal H}$, is a pure state of $S$, and each pure
state of $S$ is of this form.
\end{cqmprinciple}
We call the quantum mechanics that
retains all the old principles except SQM Principle \ref{qmprinc01}, and that follows our new
principle CQM Principle \ref{cqmprinc}, `completed quantum mechanics' and refer to it
by CQM.

The first argument for our proposal of this solution comes from earlier work in relation with the
violation of Bell inequalities by means of macroscopic entities (Aerts, 1991a). There we
introduce a macroscopic material entity that entails EPR correlations. Let us briefly describe this
entity again to state our point. 

First we represent the spin of a spin ${1 \over 2}$
quantum entity by means of the elastic sphere model that we have introduced on several occasions
(Aerts, 1986, 1987, 1991a,b, 1993, 1995, 1999a,b), and that we have called the `quantum machine'. It
is well known that the states, ray states as well as density states, of the spin of a spin ${1
\over 2}$ entity can be represented by the points of a sphere $B$ in three dimensional
Euclidean space with radius
$1$ and center
$0$. Let us denote the state corresponding to the point $w \in B$ by $p_w$. To make the
representation explicit we remark that each vector
$w
\in B$ can uniquely be written as a convex linear combination of two vectors $v =
(\sin\theta\cos\phi, \sin\theta\sin\phi, \cos\theta)$ and $-v$ on the surface of the sphere
(Fig.\ 2) i.e. $w = a \cdot v - b \cdot v = (a-b)
\cdot v$ with $a , b \in [0,1]$ and $a + b = 1$. In this way we make correspond with $w$ the density
operator
$W(w)$:
\begin{equation}
W(w) = \left( \begin{array}{cc} 
acos^2{\theta \over 2} + bsin^2{\theta \over 2} &  (a-b)sin{\theta
\over 2}cos{\theta \over 2}e^{-i\phi}  \\ (a-b)sin{\theta \over
2}cos{\theta \over 2}e^{i\phi} & asin^2{\theta \over 2} + bcos^2{\theta
\over 2} 
\end{array} \right)
\end{equation}
Each density operator can be written in this form and hence the inverse correspondence is also
made explicit. We remark that the ray states, namely the density operators that are projections,
correspond to the points on the surface of $B$.

It is much less known
that the experiments on the spin of a spin ${1 \over 2}$ quantum entity can be represented
within the same picture. Let us denote the direction in which the spin will be measured by
the diametrically opposed vectors $u$ and $-u$ of the surface of $B$ (Fig.\ 1),
and let us consider $u$ as the $z$ direction of the standard spin representation (this does
not restrict the generality of our calculation). In this case, in SQM, the spin measurement along
$u$, which we denote $e_u$, is represented by the self adjoint operator
$S = {1 \over 2} E_1 - {1 \over 2} E_2$ with 
\begin{equation}
E_1 = \left(\begin{array}{cc}
1 & 0 \\
0 & 0
\end{array} \right)
\quad E_2 = \left(\begin{array}{cc}
0 & 0 \\
0 & 1
\end{array} \right)
\end{equation}
being the spectral projections. The
SQM transition probabilities, $\mu(e_u, p_w, up)$, the probability for spin up outcome if the
state is $p_w$, and
$\mu(e_u, p_w, down)$, the probability for spin down outcome if the state is $p_w$, are then:
\begin{equation} \label{probquant}
\begin{array}{l}
\mu(e_u, p_w, up) = tr(W(w) \cdot E_1) = acos^2{\theta \over 2} + bsin^2{\theta  \over 2} \\
\mu(e_u, p_w, down) = tr(W(w) \cdot E_2) = asin^2{\theta \over 2} + bcos^2{\theta  \over 2}
\end{array}
\end{equation}
Let us now show that, using the sphere picture, we can propose a realizable mechanistic procedure
that gives rise to the same probabilities and can therefore represent the spin measurement.

\vskip 0.5 cm

\hskip 2 cm \includegraphics{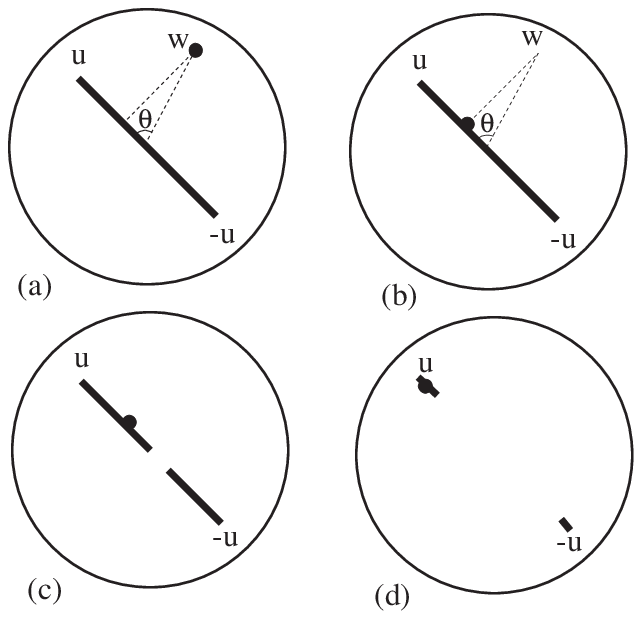}

\begin{quotation}
\noindent \baselineskip= 7pt \smallroman Fig.\ 1 : A representation of the
quantum machine. In (a) the particle is in
state $\scriptstyle p_w$, and
the elastic corresponding to the experiment $\scriptstyle
e_u$ is installed between the two diametrically opposed
points $\scriptstyle u$ and
$\scriptstyle -u$. In (b) the particle falls orthogonally onto the elastic  and sticks to
it. In (c) the elastic breaks and the particle is
pulled  towards the point
$\scriptstyle u$, such that (d) it arrives at the
point
$\scriptstyle u$, and the experiment $\scriptstyle
e_u$ gets the outcome
$\scriptstyle {\rm {\it `up'}}$.
\end{quotation}
Our mechanistic procedure starts by installing an elastic
strip (e.g. a rubber band) of 2 units of length, such that it is
fixed with one of its end-points in
$u$ and the other end-point in
$-u$ (Fig.\ 1,a).
Once the elastic is installed, the particle falls from its original place $w$ orthogonally onto
the elastic, and sticks to it (Fig.\ 1,b). Then, the elastic breaks in some arbitrary point.
Consequently the particle, attached to one of the two pieces of the elastic (Fig.\ 1,c), is
pulled to one of the two end-points $u$ or $-u$  (Fig.\ 1,d). Now, depending on whether the
particle arrives in $u$ (as in  Fig.\ 1) or in $-u$, we give the outcome {\it `up'} or {\it `down'}
to this experiment $e_u$.

Let us prove that the transition
probabilities are the same as those calculated by SQM. The probability
$\mu(e_u, p_w, up)$, that the particle ends up in point $u$ (experiment $e_u$ gives outcome
{\it `up'}) is given by the length of the piece of elastic
$L_1$ divided by the total length of the elastic. The probability,  
$\mu(e_u, p_w, down)$, that the particle ends up in point $-u$ (experiment $e_u$ gives
outcome {\it `down'}) is given by  the length of the piece of elastic $L_2$ divided by the total
length of the elastic. This means that we have:
\begin{equation} \label{probform}
\begin{array}{l}
\mu(e_u, p_w, up) = {L_1\over 2} = {1 \over 2}(1 + (a-b) 
cos\theta) = a cos^2{\theta\over 2} + b sin^2{\theta\over 2} \\
\mu(e_u, p_w, down) = {L_2\over 2} = {1 \over 2}(1 - (a-b) 
cos\theta) = a sin^2{\theta\over 2} + b cos^2{\theta\over 2}
\end{array}
\end{equation}
Comparing (\ref{probquant}) and (\ref{probform}) we see that our mechanistic procedure represents
the quantum mechanical measurement of the spin.

\vskip 0.5 cm

\hskip 3.5 cm \includegraphics{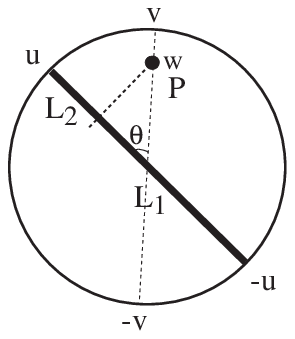}

\begin{quotation}
\noindent \baselineskip= 7pt \smallroman Fig.\ 2 : A representation of the 
experimental process.
The elastic of length 2, corresponding to the experiment
$\scriptstyle e_u$, is installed between $\scriptstyle u$ and 
$\scriptstyle -u$. The probability, $\scriptstyle \mu(e_u, p_w,
up)$, that the particle ends in point
$\scriptstyle u$ under influence of the experiment $\scriptstyle e_u$
is given by the length of the piece of elastic $\scriptstyle L_1$
divided by the total length of the elastic. The probability,
$\scriptstyle
\mu(e_u, p_w, down)$, that the particle ends  in
point $\scriptstyle -u$ is given by the length of the piece of elastic
$\scriptstyle L_2$ divided by the total length of the elastic.
\end{quotation}
To realize the macroscopic model with EPR correlations we consider two such `quantum machine'
spin models, where the point particles are connected by a rigid rod, which introduces the
correlation. We will only describe the situation where we realize a state that is
equivalent to the singlet spin state $p_s$, where $s = u_1 \otimes u_2 - u_2 \otimes u_1$, and refer
to Aerts (1991a) for a more detailed analysis. Suppose that the particles are in states
$p_{w_1}$ and
$p_{w_1}$ where
$w_1$ and
$w_2$ are respectively the centers of the spheres $B_1$ and $B_2$
(Fig.\ 3) connected by a rigid rod. We call this state (the presence of the rod included) $p_w$. The
experiment
$e_{({u_1},{u_2})}$ consists of performing $e_{u_1}$ in $B_1$ and
$e_{u_2}$ in $B_2$ and collecting the outcomes {\it (up, up), (up, down), (down, up)}
or {\it (down, down)}.
In Fig.\ 3 we show the different phases of the experiment. We make the hypothesis that one of
the elastics breaks first and pulls one of the particles {\it up} or {\it down}. Then we also make
the hypothesis that once that one of the particles has reached one of the outcomes, the rigid
connection breaks down. The experiment continues then without connection in the sphere where the
elastic is not yet broken. The joint probabilities can now easily be calculated:
\begin{equation}
\begin{array}{ll}
\mu(e_{({u_1},{u_2})}, p_w, (up,up)) = {1 \over 2} \sin^2{\alpha \over 2} &
\mu(e_{({u_1},{u_2})}, p_w, (up,down)) = {1 \over 2} \cos^2{\alpha \over 2} \\
\mu(e_{({u_1},{u_2})}, p_w, (down,up)) = {1 \over 2} \cos^2{\alpha \over 2} &
\mu(e_{({u_1},{u_2})}, p_w, (down,down)) = {1 \over 2} \sin^2{\alpha \over 2}
\end{array}
\end{equation}
where $\alpha$ is the angle between $u_1$ and $u_2$. These are exactly the quantum probabilities
when $p_w$ represents the singlet spin state. As a consequence our model is a representation of
the singlet spin state. This means that we can put $p_s = p_w$.

\vskip 0.5 cm

\hskip 0.5 cm \includegraphics{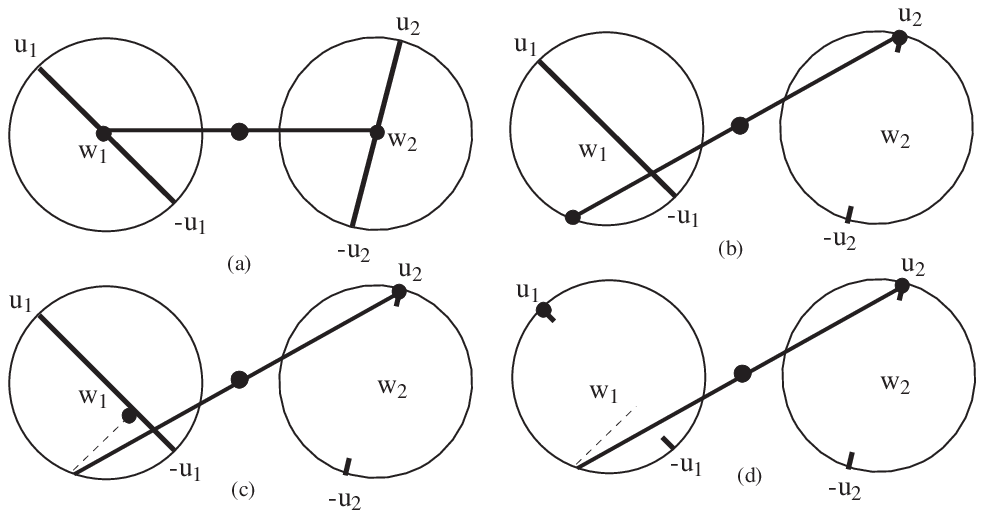}

\begin{quotation}
\hskip 0.7 cm \noindent \baselineskip= 7pt \smallroman Fig.\ 3 : A macroscopic mechanical entity with EPR
correlations.
\end{quotation}
Why does this example inspire us to put forward the hypothesis that density states are pure
states? Well, if we consider the singlet spin state, then this is obviously a nonproduct state, and
hence the states of the subentities are density states. In fact they are the density states
$p_{W_1}$ and $p_{W_2}$ where
\begin{equation}
W_1 = W_2 = \left( \begin{array}{cc}
{1 \over 2} & 0 \\
0 & {1 \over 2}
\end{array} \right)
\end{equation}
However, the state of the joint entity is clearly not given by the density state corresponding to
the density operator 
\begin{equation}
\left(
\begin{array}{cc} {1 \over 2} & 0 \\
0 & {1 \over 2}
\end{array} \right) \otimes \left( \begin{array}{cc}
{1 \over 2} & 0 \\
0 & {1 \over 2}
\end{array} \right)
\end{equation}
because this state does not entail correlations. It is due the presence of the EPR correlations
that the state of the joint entity is represented by a ray state. In our macroscopic
mechanistic model however all the states (also the states of the subentities) are `pure states' and
not mixtures (remark that we use the concept `pure state' as defined in section \ref{sec01}).
If our proposal were true, namely if density states as well as ray states in principle
represented pure states, we could also understand why, although the state of the joint entity
uniquely determines the states of the subentities, and hence Physical Principle
\ref{physprinc02} is satisfied, the inverse is not true: the states of the subentities do not
determine the state of the joint entity. Indeed, a state of one subentity cannot contain the
information about the presence of an eventual correlation between the subentities. This way, it
is natural that different types of correlations can give rise to different states of the joint
entity, the subentities being in the same states. This possibility expresses the
philosophical principle that the whole is greater than the sum of its parts and as our model shows
it is also true in the macroscopic world.

Let us now say some words about the generality of the construction that inspired us for the
proposed solution. It has been shown in Coecke (1995a) that a quantum machine like model
can be realized for higher dimensional quantum entities. Coecke (1995b, 1996) also
showed that all the states of the tensor product can be realized by introducing correlations on
the different component states. This means that we can recover all the
nonproduct ray states of the tensor product Hilbert space by identifying them with a product state
plus a specific correlation for a general quantum entity and hence that our solution of the
paradox is a possible solution for a general quantum entity.

\section{Experimental Testing of the Solution}
If we carefully analyze the calculations that show the
equivalence of our model to the quantum model, we can understand why the distinction between
`interpreting density states as mixtures' and `interpreting density states as pure states' cannot
be made experimentally. Indeed, because of the linearity of the trace,
used to calculate the quantum transition probabilities, and because the inner points of the sphere
can be written as convex linear combinations of the surface points, an ontological situation of
mixtures must give the same experimental results as an ontological situation of pure states.

If we would be able to realize experimentally a nonlinear evolution of one of the subentities that
has been brought into an entangled state with the other subentity as subentity of a joint entity,
it would be possible to test our hypothesis and to detect
experimentally whether density states are pure states or mixtures. Indeed, suppose that we a
nonlinear evolution of one of the entangled subentities could be realized. Then, we can
distinguish the two possibilities in the following way. If the density state $p_{W_1}$ of the
entangled subentity is a mixture, then this state evolves while staying a convex linear combination
of the ray states $p_{v_1}$ and $p_{-v_1}$ (referring to the situation of Fig.\ 3). The nonlinear
evolution makes evolve the ray states $p_{v_1}$ and $p_{-v_1}$ and this determines
the evolution of the density state $p_{W_1}$, but the correspondence between $p_{W_1}$ and
$p_{v_1}$ and $p_{-v_1}$ remains linear. If the density state $p_{W_1}$ of the entangled subentity
is a pure state, then the nonlinear evolution will make it evolve independent of the way in which
the ray states $p_{v_1}$ and $p_{-v_1}$ evolve. This means that in general the relation between
$p_{W_1}$ and $p_{v_1}$ and $p_{-v_1}$ will not remain that of a convex linear combination. So we
can conclude that for a nonlinear evolution the change of the density state of an entangled
subentity under this evolution will be different depending on whether it is a mixture or a pure
state. This difference can experimentally be detected by a proper experimental setup. We believe
that such an experiment would be of great importance for the problem that we have outlined here.

\section{Consequences for Quantum Axiomatics}
The quantum axiomatic approaches make use of Piron's representation theorem
where the set of pure states is represented as the set of rays of a generalized Hilbert space
(Piron, 1964, 1976). This theorem has meanwhile been elaborated and the recent result of Sol\`er has
made it possible to formulate an axiomatics that characterizes SQM for real, complex or
quaternionic Hilbert spaces (S\`oler 1995, Aerts and Van Steirteghem 1999). This standard axiomatic
approach aims to represent pure states by rays of the Hilbert space. If our proposal is true, an
axiomatic system should be constructed that aims at representing pure states by means of density
operators of the Hilbert space. Within the generalization of the Geneva-Brussels approach that we
have formulated recently, and where the mathematical category is that of state property systems
and their morphisms, such an axiomatic can be developed (Aerts, 1999a; Aerts et al. 1999; Van
Steirteghem, 1999; Van der Voorde 1999). In Aerts (1999c) we made a small step in the direction of
developing such an axiomatic system by introducing the concept of `atomic pure states' and treating
them as earlier the pure states were treated, aiming to represent these atomic pure states by the
rays of a Hilbert space. We proved that in this case the covering law remains a problematic axiom in
relation to the description of the joint entity of two subentities (Theorem 18 of Aerts 1999c).
We are convinced that we would gain more understanding in the joint entity problem if a new
axiomatic would be worked out, aiming to represent pure states by density operators of a Hilbert
space, and we are planning to engage in such a project in the coming years.

\section{Acknowledgments}
Diederik Aerts is senior research associate of the Belgian Fund for Scientific Research. This
research was carried out within the project BIL 96-03 of the Flemish Governement.

\section{References}
Aerts, D., (1982). Description of Many Physical Entities without the Paradoxes Encountered in
Quantum Mechanics, {\it Found. Phys.}, {\bf 12}, 1131-1170.

\smallskip
\noindent
Aerts, D., (1984a).  Construction of the Tensor Product for Lattices of Properties of
Physical Entities, {\it J. Math. Phys.}, {\bf 25}, 1434-1441.

\smallskip
\noindent
Aerts, D., (1984b). The Missing Elements of Reality in the Description of Quantum Mechanics of the
EPR Paradox Situation, {\it Helv. Phys. Acta}, {\bf 57}, 421-428.

\smallskip
\noindent
Aerts, D., (1984c). How Do We Have to Change Quantum Mechanics in Order to Describe Separated
Systems, in {\it The Wave-Particle Dualism}, eds. S. Diner et al., D. Reidel Publishing
Company, 419-431.

\smallskip
\noindent
Aerts, D., (1985a). The Physical Origin of the Einstein Podolsky Rosen Paradox, in {\it Open
Questions in Quantum Physics}, eds. G. Tarozzi and A. van der Merwe, D. Reidel Publishing Company,
33-50. 

\smallskip
\noindent
Aerts, D., (1985b). The Physical Origin of the EPR Paradox and How to Violate Bell Inequalities by
Macroscopic Systems, in {\it On the Foundations of Modern Physics}, eds. P. Lathi and P.
Mittelstaedt, World Scientific, Singapore, 305 - 320.

\smallskip
\noindent
Aerts, D., (1986). A Possible Explanation for the Probabilities of Quantum Mechanics,
{\it J. Math. Phys.}, {\bf 27}, 202-210.

\smallskip
\noindent
Aerts, D., (1987). The Origin of the Nonclassical Character of the Quantum Probability
Model, in {\it Information, Complexity, and Control in Quantum Physics}, eds. A.
Blanquiere, S. Diner, and G. Lochak, Springer-Verlag, Wien-New York, 77 - 100. 

\smallskip
\noindent
Aerts, D., (1991a). A Mechanistic Classical Laboratory Situation Violating the Bell
Inequalities with $\sqrt 2$, Exactly 'in the Same Way' as its Violations by the EPR
Experiments, {\it Helv. Phys. Acta}, {\bf 64}, 1 - 24.

\smallskip
\noindent
Aerts, D., (1991b). A Macroscopic Classical Laboratory Situation with Only Macroscopic
Classical Entities Giving Rise to a Quantum Mechanical Probability Model, in
{\it Quantum Probability and Related Topics, Volume VI}, ed. L. Accardi, World Scientific
Publishing, Singapore, 75 - 85.

\smallskip
\noindent
Aerts, D., (1993). Quantum Structures Due to Fluctuations of the Measurement Situations,
{\it Int. J. Theor. Phys.}, {\bf 32}, 2207 - 2220.

\smallskip
\noindent
Aerts, D., (1995). Quantum Structures:  An Attempt to Explain their Appearance in
Nature, {\it Int. J. Theor. Phys.}, {\bf 34}, 1165 - 1186.

\smallskip
\noindent
Aerts, D., (1999a). Foundations of Quantum Physics: a General Realistic and Operational
Approach, {\it Int. J.  Theor. Phys.}, {\bf 38}, 289 - 358.

\smallskip
\noindent
Aerts, D., (1999b). The Stuff the World is Made Of: Physics and Reality, to
appear in {\it Einstein meets Magritte: An Interdisciplinary Reflection}, eds. D. Aerts, J.
Broekaert and E. Mathijs, Kluwer Academic, Dordrecht, Boston, London.

\smallskip
\noindent
Aerts, D., Colebunders, E., Van der Voorde, A. and Van Steirteghem, B., (1999). State Property
Systems and Closure Spaces: A Study of Categorical Equivalence, {\it Int. J. Theor. Phys.}, {\bf
38}, 359-385.

\smallskip
\noindent
Aerts, D., Broekaert, J. and Smets, S. (1999). Inconsistencies in
Constituent Theories of World Views : Quantum Mechanical Examples, {\it Foundations of Science},
{\bf 3}, to appear.

\smallskip
\noindent
Aerts, D., Coecke, B., D'Hooghe, B., Durt, T. and Valckenborgh, F., (1996). A Model with
Varying Fluctuations in the Measurement Context, in {\it Fundamental Problems in Quantum
Physics II}, eds. Ferrero, M. and van der Merwe, A., Plenum, New York, 7-9.

\smallskip
\noindent
Aerts, D. and Valckenborgh, F., (1999). Lattice Extensions and the Description of
Compound Entities, in preparation.

\smallskip
\noindent
Aerts, D. and Van Steirteghem, B., (1999). Quantum Axiomatics and a Theorem of M. P. Sol\`er,
submitted to {\it International Journal of Theoretical Physics}.

\smallskip
\noindent
Bell, J., (1964). On the Einstein Podolsky Rosen Paradox, {\it Physics 1}, {\bf 3}, 195.

\smallskip
\noindent
Coecke, B., (1995a). Representation for Pure and Mixed States of Quantum Physics in Euclidean
Space, {\it Int. J. Theor. Phys.}, {\it 34}, 1165.

\smallskip
\noindent
Coecke, B., (1995b). Representation of a Spin-1 Entity as a Joint System of Two Spin-1/2 Entities
on which we Introduce Correlations of the Second Kind, {\it Helv. Phys. Acta}, {\bf 68}, 396.

\smallskip
\noindent
Coecke, B. (1996). Superposition States through Correlations of the Second Kind, {\it Int. J. Theor.
Phys.}, {\bf 35}, 1217.

\smallskip
\noindent
Einstein, A., Podolsky, B. and Rosen, N., (1935). Can Quantum Mechanical Reality
considered to be Complete? {\it Phys. Rev.} {\bf 47}, 777.

\smallskip
\noindent
Piron, C., (1964). Axiomatique Quantique, {\it Helv. Phys. Acta}, {\bf 37}, 439.

\smallskip
\noindent
Piron, C., (1976). {\it Foundations of Quantum Physics}, Benjamin, New York.

\smallskip
\noindent
Pulmannov\`a, S., (1983). Coupling of Quantum Logics, {\it Int. J. Theor. Phys.}, {\bf 22}, 837.

\smallskip
\noindent
Pulmannov\`a, S., (1985). Tensor Product of Quantum Logics, {\it J. Math. Phys.}, {\bf 26}, 1.

\smallskip
\noindent
Randall, C. and Foulis, D., (1981). Operational Statistics and Tensor
Products, in   {\it Interpretations and Foundations of Quantum Theory}, H.
Neumann, ed., B.I. Wissenschaftsverslag, Bibliographisches Institut,
Mannheim, 21.

\smallskip
\noindent
Sol\`er M.P., (1995). Characterization of Hilbert Spaces by Orthomodular Spaces, {\it Comm.
Algebra}, {\bf 23}, 219-243.

\smallskip
\noindent
Valckenborgh, F., (1999). Structures for the Description of Compound Physical Systems, submitted
to {\it International Journal of Theoretical Physics}.

\smallskip
\noindent
Van Fraassen, B. C., (1991). {\it Quantum Mechanics: An Empiricist View}, Oxford University
Press, Oxford, New York, Toronto.

\smallskip
\noindent
Van der Voorde, A., (1999). A Categorical Approach to T$_1$ Separation and the Product of State
Property Systems, submitted to {\it International Journal of Theoretical Physics}.

\smallskip
\noindent
Van Steirteghem, B., (1999). T$_0$ Separation in Axiomatic Quantum Mechanics, submitted to {\it
International Journal of Theoretical Physics}.

\smallskip
\noindent
von Neumann, J., (1932). {\it Mathematische Grundlagen der Quantenmechanik},
Springer-Verlag, Berlin.

\end{document}